\documentclass[12pt]{article}
\begin{document}
\title{EFFECT OF MUONS ON THE PHASE TRANSITION IN MAGNETISED PROTO-NEUTRON STAR MATTER}
\author{Asha Gupta, V.K.Gupta \footnote{E--mail : vkg@ducos.ernet.in}, S.Singh\footnote{E--mail : santokh@ducos.ernet.in}, J.D.Anand   \\
 	{\em Department of Physics and Astrophysics,} \\
	{\em University of Delhi, Delhi-110 007, India.} \\
        {\em Inter-University Centre for Astronomy and Astrophysics,} \\
        {\em Ganeshkhind, Pune 411007 , India.} \\
        }
\renewcommand{\today}{}
\setlength\textwidth{5.75 in}
\setlength\topmargin{-1.cm}
\setlength\textheight{8 in}
\addtolength\evensidemargin{-1.cm}
\addtolength\oddsidemargin{-1.cm}
\font\tenrm=cmr10
\def\baselinestretch{1.4}

\maketitle
\large
\begin{abstract}
We study the effect of inclusion of muons and the muon neutrinos on the phase transition from nuclear to quark matter in a magnetised proto-neutron star and compare our results with those obtained by us without the muons. We find that the inclusion of muons changes slightly the nuclear density at which transition occurs.However the dependence of this transition density on various chemical potentials, temperature and the magnetic field remains quantitatively the same.   
\end{abstract}
\pagebreak
\begin{section} {Introduction}

The phase transition from hadron matter to quark matter (QM) is expected to occur in a variety of physically different situations. In the astrophysical context this transition can take place in the interior of a cold, 
extremely dense neutron star. The other and the more likely situation in which such a transition can take place is the proto-neutron star \cite{prakash,ellis}.  
These proto-neutron stars (PNS) have very high temperatures in their interiors ($\sim$50MeV) and more importantly, have a large amount of trapped neutrinos \cite{burrows,bethe,keil}, both electron type and muon type. This PNS stage lasts only for a few seconds during which time most of its energy and neutrinos are released. In one of the recent studies Lugones and Benvenuto \cite{lugones} have considered the transition from nuclear matter (NM) to QM under conditions prevailing in a PNS during the first tens of seconds of its birth. They considered a hyperonic equation of state (EOS) for the NM phase and the `bag model' EOS for the QM phase. They studied the effect of the mass of the strange quark, $m_s$, and the bag constant, B, on the phase transition.\\
Another feature of the neutron stars, proto-neutron stars and other such dense stellar objects is the presence of intense magnetic fields which may go upto as high as $\sim$ 8$\times10^{14}$ G (Kuoveliotou $\b{et}$ $\b{al}$ \cite{kuo}). 
A large number of studies have been made of the effect of magnetic field on the NM as well as on the QM EOS at various densities and for various models \cite{deb,som,ben,anand,gupta,ashok,in}. Infact some of the studies have gone upto unrealistically high magnetic fields ($>10^{20}$ G) \cite{deb,som}. Many investigations have also been made of the effect of magnetic field on the mass-radius relationship, as well as the frequencies of radial modes of both rotating and non-rotating stars.\\
In a preliminary investigation we have recently studied the effect of magnetic field on the phase transition on PNS \cite{guptaa}. In this investigation we had included the electron and the corresponding electron neutrinos only. We found that the magnetic field has a significant effect on the density of nuclear matter at which the transition to quark matter can take place. Since the muon has a mass nearly 200 times that of electron, the effect of the magnetic field is expected to change significantly in the presence of the muons. Motivated by this we now include the muon and of course the muon neutrino in our investigation. We have gone upto a magnetic field of 3$\times10^4$ $MeV^2$ (1$~MeV^2$ $\sim$ 1.6$\times10^{14}$ G). This is about the maximum field allowed by the virial theorem. Also Agasian $\b{et}$ $\b{al}$ \cite{agasian} have studied the effect of a strong magnetic field and reached a tentative conclusion that the gluon condensate could break down at fields above (0.5 - 1)$\times10^{19}$ G leading to dramatic consequences for an effective QCD model like the present one or the bag model for that matter.\\ 
  
\end{section}

\vspace {0.5cm}
\begin{section} {The transition from nuclear matter to quark matter} 

We follow here the same approach and criterion for phase transition as in our earlier work \cite{guptaa}. In particular we have used the relativistic mean field theoretic model for the hadron sector, the density dependent quark mass model for the quark sector and taken the NM to QM phase transition to be a first order phase transition. The inclusion of muons and their neutrinos leads only to minor modifications in the formalism. In the expressions for the total pressure P and the mass energy density $\rho$:  
\begin{eqnarray}
P &=& \sum_{i} P_i + \frac{1}{2}\left(\frac{g_\omega}{m_\omega}\right)^2~ \rho_{B}^{'2} - \frac{1}{2}\left(\frac{g_\sigma}{m_\sigma}\right)^{-2} (g_\sigma~\sigma)^2 - \frac{1}{3}~b~m_N~(g_\sigma~\sigma)^3 \nonumber \\ 
&-& \frac{1}{4}~c~(g_\sigma~\sigma)^4 + \frac{1}{2}\left(\frac{g_\rho}{m_\rho}\right)^2~\rho_{I_3}^{'2} 
\end{eqnarray}
\begin{eqnarray}
\rho &=& \sum_{i} \rho_i + \frac{1}{2}\left(\frac{g_\omega}{m_\omega}\right)^2~ \rho_{B}^{'2} + \frac{1}{2}\left(\frac{g_\sigma}{m_\sigma}\right)^{-2} (g_\sigma~\sigma)^2 + \frac{1}{3}~b~m_N~(g_\sigma~\sigma)^3 \nonumber \\ 
&+& \frac{1}{4}~c~(g_\sigma~\sigma)^4 + \frac{1}{2}\left(\frac{g_\rho}{m_\rho}\right)^2~\rho_{I_3}^{'2} 
\end{eqnarray}
the summation over i will now include the muons and their neutrinos as well. The expressions for the energy, pressure and number density of $\mu$ and $\nu_\mu$ are respectively the same as for e and $\nu_e$ given in \cite{guptaa}, for both the non-magnetic and magnetic case. Similarly in the presence of muons, charge neutrality condition is modified to 
\begin{equation}
n_p = n_e + n_\mu
\end{equation}
and the $\beta$-equilibrium condition to
\begin{equation}
\mu_n- \mu_p = \mu_e - \mu_{\nu_e}=\mu_{\mu}-\mu_{\nu_\mu}
\end{equation}
In order to compute the conditions for phase transition, we apply the standard Gibb's criteria i.e, equality of pressure, P, temperature, T and Gibb's energy per baryon, g, in both phases: 
\begin{equation}
P_q = P_h ~,       ~~~  T_q=T_h ~,  ~~~   g_q=g_{h}
\end{equation}
together with flavour per baryon conservation equations
\begin{equation}
Y_i^q = Y_i^h
\end{equation}
for i=u, d, e, $\mu$, $\nu_e$, $\nu_\mu$. Notice that this condition automatically makes the quark phase to be charge neutral. Here
\begin{equation} 
Y_i^x = \frac{n_i^x}{n_B}             
\end{equation}
for  x =  h or q. Also
\begin{eqnarray}
Y_u^h & = & 2Y_p +Y_n \nonumber \\
Y_d^h & = & Y_p + 2Y_n
\end{eqnarray}
Starting with $Y_p$ and $Y_n$, obtained from the hadronic phase, one obtains various number densities and thereby the chemical potentials in the quark phase; energy and pressure are then obtained from their respective expressions \cite{guptaa}.
\end{section}

\vspace {0.5cm}
\begin{section} { Results and discussion } 
In this work we are interested in the deconfinement transition from hadron to quark matter in the presence of magnetic field and trapped neutrinos of both type. The details of the calculations are given in \cite{guptaa}. For a fixed set of values of $\mu_{\nu_e}$, $\mu_{\nu_\mu}$, T and B the nuclear phase depends only on one of the chemical potentials (say $\mu_e$). With $\mu_e$, $\mu_{\nu_e}$, $\mu_{\nu_\mu}$, T and B fixed, we can calculate various number densities and the corresponding abundances. Using conditions (5), (6) we obtain the mass energy density, U, of hadron matter at which it deconfines.
At finite temperatures, we have to evaluate numerically Fermi-Dirac integrals appearing in expressions for $n_i$, $n_{i}^{s}$, $P_i$ and $E_i$. The procedure for numerical integrals originally due to Cloutman \cite{law} has been discussed in detail in our earlier work \cite{guptaa} and need not be elaborated in detail here. Throughout we have fixed the DDQM parameter C at 90 MeV$fm^{-3}$.\\
\\
In fig 1, we have plotted the mass energy density, U, of hadron matter (in terms of the nuclear saturation density, $\rho_o$ = 2.7 $\times 10^{14}$gm$cm^{-3}$) at which phase transition occurs, versus $\mu_{\nu_\mu}$, the muon neutrinos chemical potential at $\mu_{\nu_e}$ = 0, T = 0, 60 MeV and B = 0, 3$\times10^4$ $MeV^2$ (1$~MeV^2$ $\sim$ 1.6$\times10^{14}$ G). The general behaviour of these curves is very similar to what we obtained in \cite{guptaa}, without the muon and the muon neutrinos. The transition energy density increases with increasing $\mu_{\nu_\mu}$ at nearly the same rate as with $\mu_{\nu_e}$ in the earlier paper. Also the transition density increases marginally with increasing magnetic field except for low temperatures (T = 10 MeV) and high $\mu_{\nu_\mu}$ ( $\geq$ 150 MeV) when the density decreases with increasing magnetic field. Figure 2 is the same as figure 1 except that $\mu_{\nu_e}$ = 200 MeV. We find that at this increased electron neutrinos chemical potential, the effect of $\mu_{\nu_\mu}$ is same in trend as for $\mu_{\nu_e}$ = 0 but significantly reduced in magnitude.\\
In figures 3 and 4 we have plotted the transition density vs the transition temperature for $\mu_{\nu_\mu}$ = 0, 100, 200 MeV, B = 0, 3$\times10^4$ $MeV^2$ for $\mu_{\nu_e}$ = 0 (figure 3) and 200 MeV (figure 4). We find a steep fall in U with increasing temperature. The fall is much steeper for $\mu_{\nu_e}$ = 0 than for $\mu_{\nu_e}$ = 200 MeV. For a given $\mu_{\nu_e}$ the fall in U is somewhat steeper for $\mu_{\nu_\mu}$ = 0 MeV than for $\mu_{\nu_\mu}$ = 100 or 200 MeV. Also beyond T $\geq$ 20 MeV the U vs T graph is almost linear particularly at high $\mu_{\nu_e}$ and $\mu_{\nu_\mu}$. In figure 5 we show the behaviour of U as a function of the magnetic field. We find that at high neutrino chemical potentials, where the transition density is high it decreases with increasing magnetic field, whereas for lower neutrino chemical potentials and high T, where the transition density is much lower,it either remains constant or increases very slowly with increasing magnetic field. The most promising scenario for the conversion of NM to QM in a PNS is where the PNS has a high temperature or low neutrino content. But in such a situation the transition is insensitive to the magnetic field, as has already been observed in \cite{guptaa}.\\
 These observations are somewhat contrary to our initial expectations. It appears that the inclusion of the $\mu$-mesons does not have a very significant effect on the parameters of phase transition, viz the energy density at which the transition is likely to take place. Also despite the fact that the muons have a mass 200 times that of electrons, the effect of the magnetic field on the system seems to be nearly the same whether the leptons present in the system are (e, $\nu_e$) or ($\mu$, $\nu_\mu$). A closer scrutiny of the result reveals that it is the total neutrino potential which is important and not the individual values of  $\mu_{\nu_e}$ and $\mu_{\nu_\mu}$. Though the muon mass is not negligible compared either to the temperatures under consideration or the relevant chemical potential or the magnetic field ( 0 $\leq$ $\sqrt{eB}$ $\leq$ $\sqrt{3}$ $\times10^{2}$ MeV), unlike the electron mass, the total contribution of the leptons to the energy density and the pressure may not be very large, so that wherther muons are included or not does not seem to make much of a difference to the phase transition. Though in principle even the rest of the octet should also be included, in the light of our results with the muons, the octet is not likely to make any significant change in the effect of the magnetic field on the transition.    
     
\end{section}

\pagebreak

Figure captions
\vskip 0.5 cm
Figure 1. $\mu_{\nu_\mu}$ vs mass energy density, U, of hadron matter (in terms of the nuclear saturation density, $\rho_o$ = 2.7 $\times 10^{14}$gm$cm^{-3}$) at which phase transition occurs at $\mu_{\nu_e}$ = 0 MeV. Labels a, b correspond to T = 10 MeV and eB = 0, 3$\times10^4$ $MeV^2$ respectively, whereas c, d correspond to T = 60 MeV and eB = 0, 3$\times10^4$ $MeV^2$ respectively.    
\vskip 0.5 cm
Figure 2. The same as fig.1 but for $\mu_{\nu_e}$ = 200 MeV.
\vskip 0.5 cm
Figure 3. The hadron matter mass energy density of phase transition, U, (in terms of $\rho_o$) vs the temperature T at $\mu_{\nu_e}$ = 0 MeV. Labels a, b correspond to $\mu_{\nu_\mu}$ = 0.0 and eB = 0, 3$\times10^4$ $MeV^2$ respectively. Labels c, d correspond to $\mu_{\nu_\mu}$ = 100 MeV and eB = 0, 3$\times10^4$ $MeV^2$ respectively. Labels e, f correspond to $\mu_{\nu_\mu}$ = 200 MeV and eB = 0, 3$\times10^4$ $MeV^2$ respectively.
\vskip 0.5 cm
Figure 4. The same as fig.3 but for $\mu_{\nu_e}$ = 200 MeV.
\vskip 0.5 cm
Figure 5. Magnetic field B vs U. The three digits in the labels refer to temperature (0, 60 MeV), $\mu_{\nu_e}$ (0, 200 MeV) and $\mu_{\nu_\mu}$ (0, 200 MeV) respectively. Also 1 and 2 refer to first and second value in the brackett. For example: label 121 corresponds to T = 10 MeV, $\mu_{\nu_e}$ = 200 MeV and $\mu_{\nu_\mu}$ = 0 MeV.  
\pagebreak

\begin{figure}[ht]
\vskip 15truecm
\includegraphics{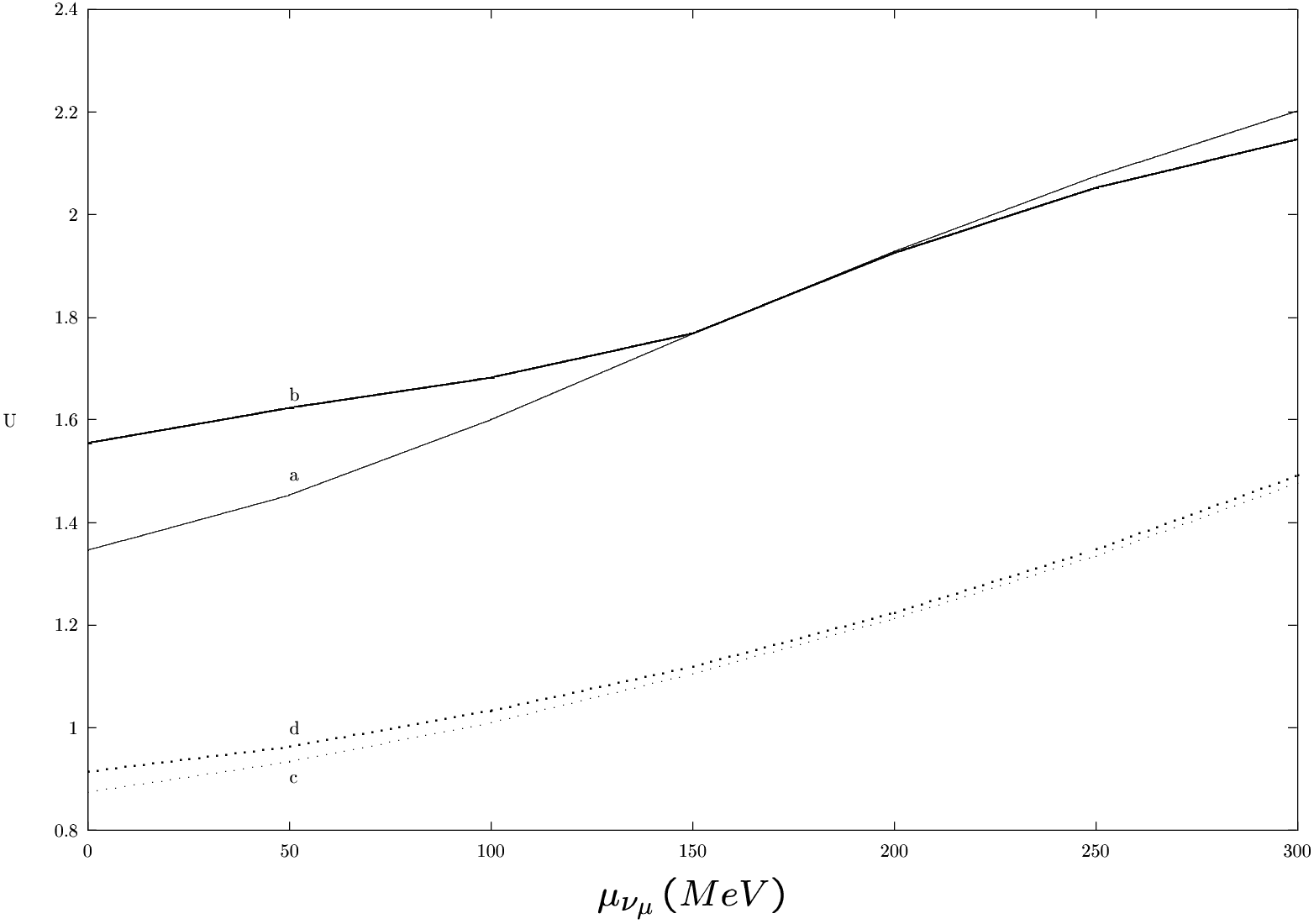}
\caption{The chemical potential of the muon neutrinos ($\mu_{\nu_\mu}$) vs mass energy density, U, of hadron matter (in terms of the nuclear saturation density, $\rho_o$ = 2.7 $\times 10^{14}$gm$cm^{-3}$) at which phase transition occurs at $\mu_{\nu_e}$ = 0 MeV. Labels a, b correspond to T = 10 MeV and eB = 0, 3$\times10^4$ $MeV^2$ respectively, whereas c, d correspond to T = 60 MeV and eB = 0, 3$\times10^4$ $MeV^2$ respectively.}
\end{figure}

\begin{figure}[ht]
\vskip 15truecm
\includegraphics{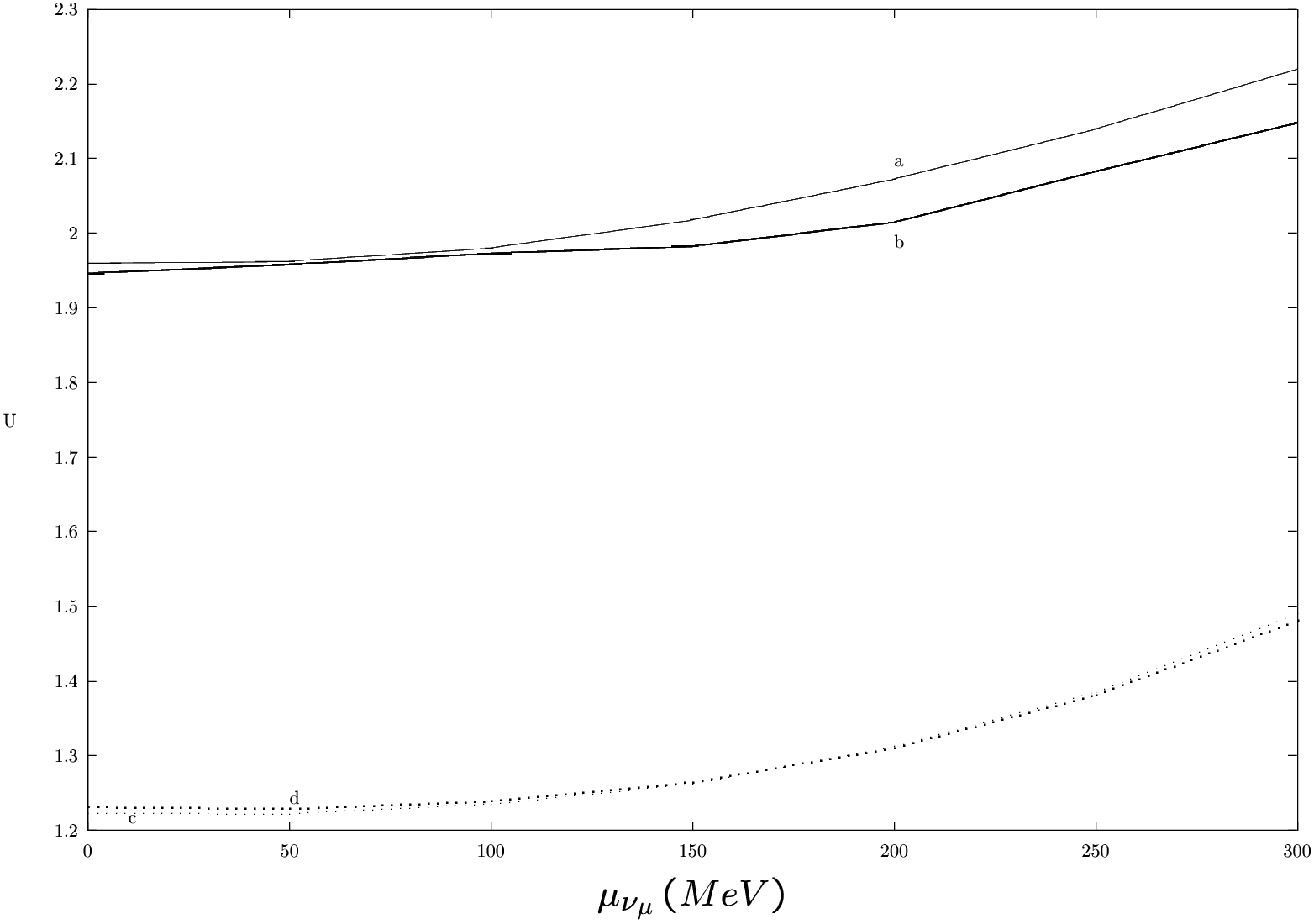}
\caption{The same as fig.1 but for $\mu_{\nu_e}$ = 200 MeV.}
\end{figure}

\begin{figure}[ht]
\vskip 15truecm
\includegraphics{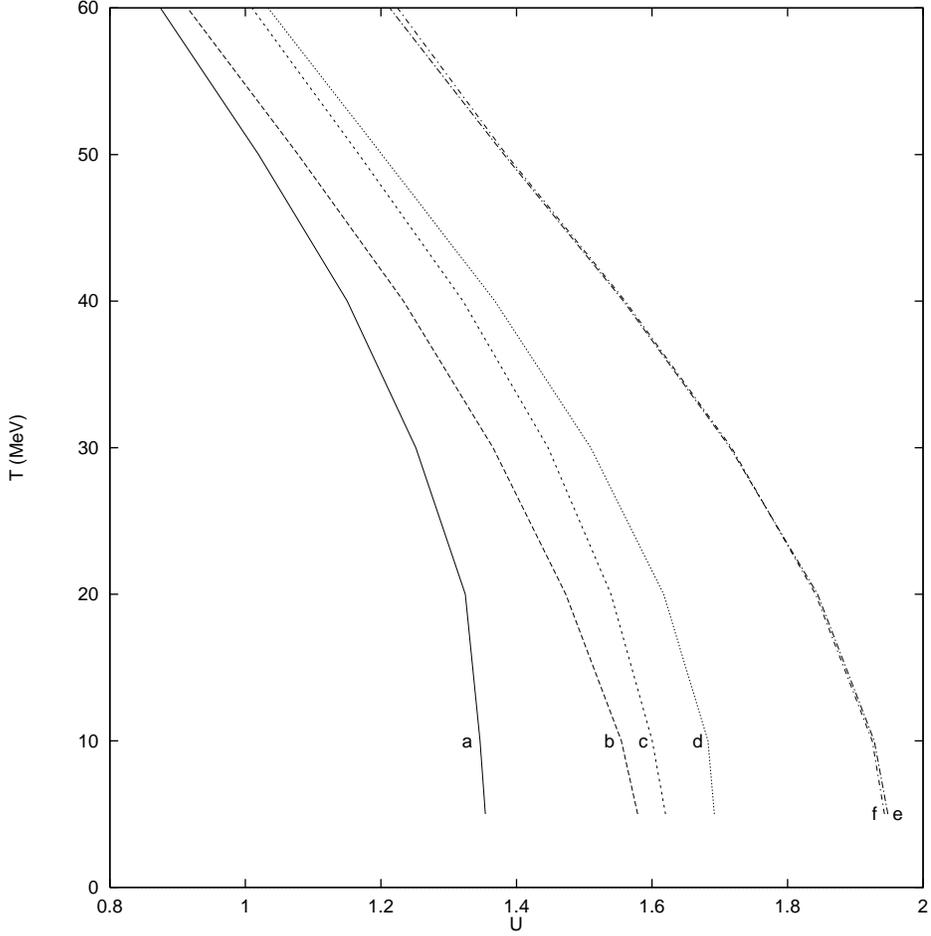}
\caption{The hadron matter mass energy density of phase transition, U, (in terms of $\rho_o$) vs the temperature T at $\mu_{\nu_e}$ = 0 MeV. Labels a, b correspond to $\mu_{\nu_\mu}$ = 0.0 and eB = 0, 3$\times10^4$ $MeV^2$ respectively. Labels c, d correspond to $\mu_{\nu_\mu}$ = 100 MeV and eB = 0, 3$\times10^4$ $MeV^2$ respectively. Labels e, f correspond to $\mu_{\nu_\mu}$ = 200 MeV and eB = 0, 3$\times10^4$ $MeV^2$ respectively.}
\end{figure}

\begin{figure}[ht]
\vskip 15truecm
\includegraphics{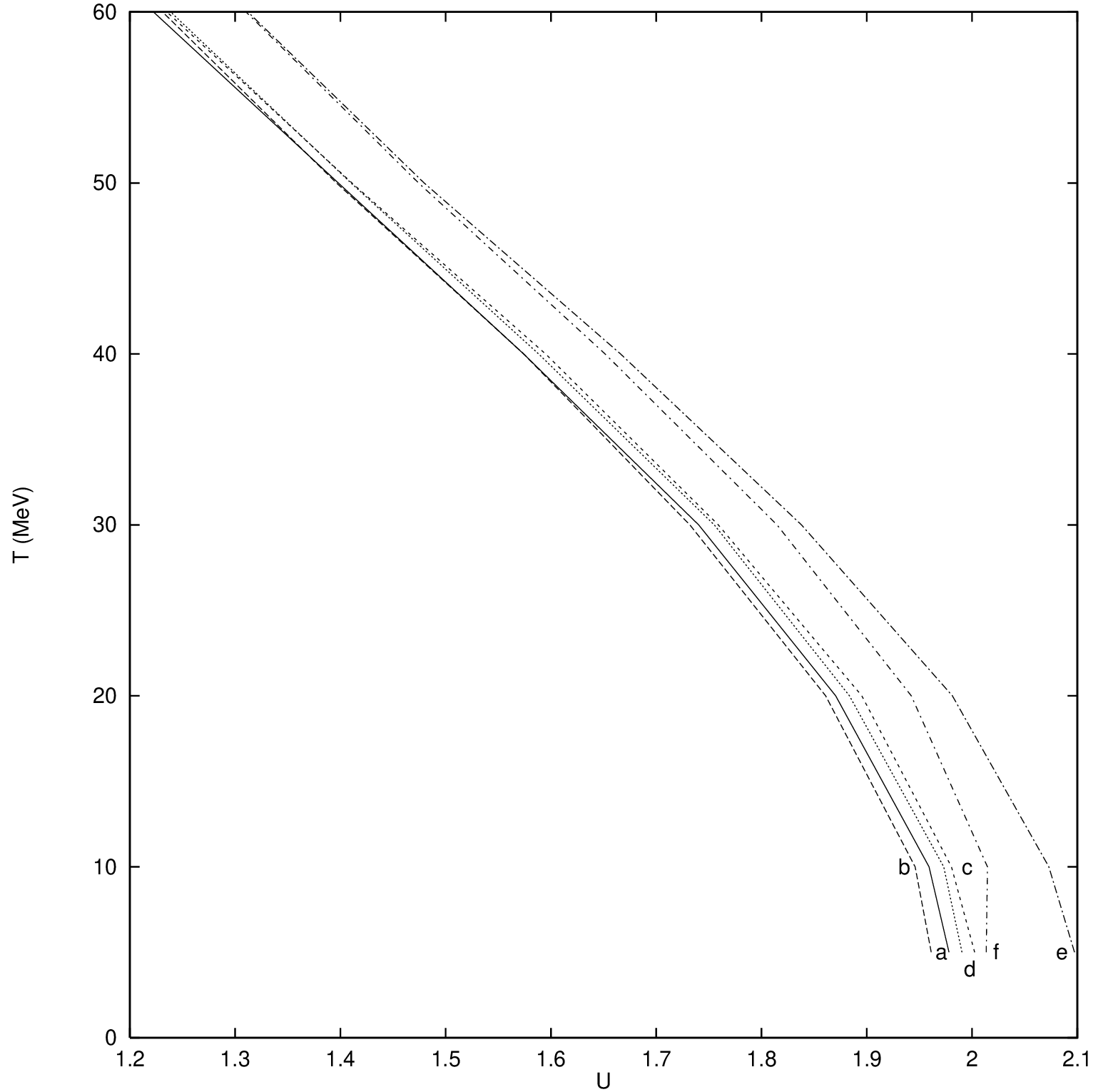}
\caption{The same as fig.3 but for $\mu_{\nu_e}$ = 200 MeV.}
\end{figure}

\begin{figure}[ht]
\vskip 15truecm
\includegraphics{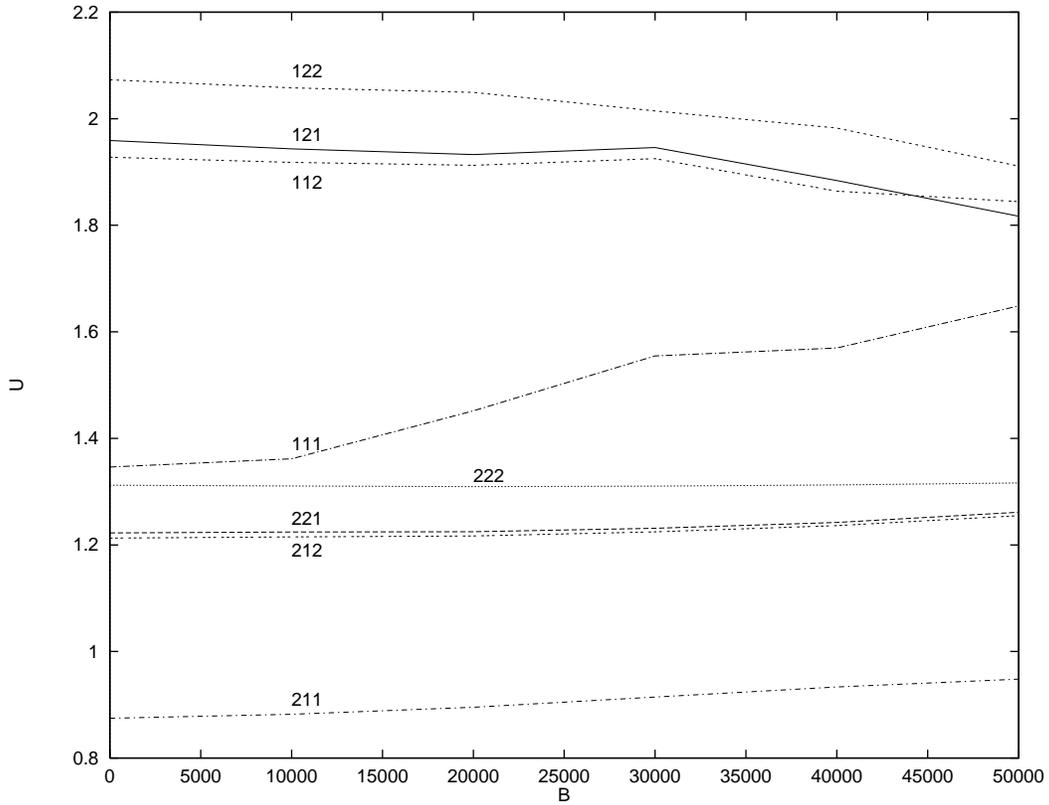}
\caption{Magnetic field B vs U. The three digits in the labels refer to temperature (0, 60 MeV), $\mu_{\nu_e}$ (0, 200 MeV) and $\mu_{\nu_\mu}$ (0, 200 MeV) respectively. Also 1 and 2 refer to first and second value in the brackett. For example: label 121 corresponds to T = 10 MeV, $\mu_{\nu_e}$ = 200 MeV and $\mu_{\nu_\mu}$ = 0 MeV.}
\end{figure}

\end{document}